\documentclass[doublecol]{epl2} 

\usepackage{graphicx}
\usepackage{subfigure}
\usepackage{amsfonts}
\usepackage{bm}
\usepackage{dcolumn}
\usepackage{color}
\usepackage{mathtools}

\usepackage{mathrsfs}
\usepackage{dsfont}
\usepackage{wasysym}

\newcommand{\bs}{\boldsymbol}

\title{The dynamics of a self-phoretic Janus swimmer near a wall}
\date{\today}

\author{Y. Ibrahim and T. B. Liverpool}

\institute{School of Mathematics, University of Bristol, Clifton, Bristol BS8 1TW, U.K. }

\abstract{
We study the effect of a nearby planar wall on the propulsion of a phoretic Janus micro-swimmer driven by asymmetric reactions on its surface which absorb reactants and generate products. We show that the behaviour of these swimmers near a wall can be classified {\bf based on whether} the swimmers are {\bf mainly} absorbing or producing reaction solutes {\bf and whether}  their swimming directions are such that the inert or active face is at the front.  We find that the wall-induced solute gradients always promote swimmer propulsion along the wall while the effect of hydrodynamics leads to  re-orientation of the swimming direction away from the wall. 
}

\begin{document} 

\maketitle

\thispagestyle{empty}
\section{Introduction}

Active materials are condensed matter systems self-driven out of equilibrium by  
components that convert stored energy into movement. They  have generated much interest  in recent years, both as inspiration for 
a new generation of smart materials and as a framework to understand 
aspects of  cell motility~\cite{MCM-RMP,Julicher:2007rx,Toner2005,Ramaswamy2010}. 
Active materials  exhibit interesting non-equilibrium phenomena, such as swarming, pattern formation and dynamic cluster formation \cite{theurkauff2012dynamic,Palacci}. 
Many of the components of active matter have come from biological systems, e.g. mixtures of cytoskeletonal polymers and motors or suspensions of  swimming micro-organisms but there has been 
an increasing interest on synthetic active components which provide promise of a variety of applications  from chemical industry to biomedical sciences~\cite{BarabanCargoDel}. 
A paradigmatic component of this type is a synthetic micro-swimmer. 
However, designing synthetic micro-scale swimmers with comparable functionality and robustness to their natural counterparts remains a challenge~\cite{paxton2006catalytically,howse2007self,ebbens2010pursuit}.   A good candidate for such synthetic micro-swimmers are self-phoretic Janus swimmers, colloidal particles with asymmetric  catalytic physico-chemical properties over their surface \cite{Golestanian01,Seifert01}.  Due to the asymmetric distribution of catalyst on their surface, they  generate or absorb chemical solutes in an asymmetric manner leading to an asymmetric distribution of solutes in the vicinity of the colloid. The coupled asymmetric distribution of the chemical solutes with the short-range solute-to-colloid surface interaction leads to the swimmer propulsion \cite{Anderson01}. Of particular importance is the behaviour of semi-dilute or concentrated  suspensions of such particles which requires an understanding and ability to predict their swimming behaviour in confinement. 

\par The first step towards understanding the behaviour of swimmers in confinement is provided by the study of their motion near planar walls. 
There have been a number of recent experiments  addressing this issue. 
A single Janus swimmer confined to a micro channel has shown a rich dynamics with the swimmer sliding along the wall while weakly rotating away from the wall. 
This reorientation continues until subsequent reflection from the wall \cite{kreuter2013transport}.  Light activated phoretic colloidal swimmers 
have been shown to swim only when close to a boundary surface \cite{Palacci}.
These suggests wall effects are a combination of wall induced  distortion {\bf both} of  fluid flow and  of the solute gradients generated by the swimmer.

\par Recent theoretical work on swimmers near walls however has tended to focus on the effect of hydrodynamic mechanisms, i.e.  on the behaviour of the fluid flow generated by swimmers near boundaries~\cite{kreuter2013transport,berke2008hydrodynamic,Stark01,Gao_Jin2014hydro,Lauga,Crowdy,Ishimoto} making the assumption that they are the dominant contributor to the motion. This is obviously the case for swimmers driven by mechanical surface distortions
~\cite{Ishimoto,Gao_Jin2014hydro}. However, it is not clear that this is also true for chemically driven swimmers for which numerical studies have shown a rich behaviour that is difficult to understand within this framework~\cite{Popescu2009cavity,Popescu2014numerics}. Here, we theoretically examine the validity of this assumption for self-phoretic swimmers near walls and seek to understand better the role of the solute gradient distortion on the dynamic behaviour of a Janus swimmer near an infinite planar wall.  We find in contrast however that  the distortion of the local gradient of solute concentration by the wall can be the dominant effect on the {\bf translational} dynamics while distortion of the fluid flow is the dominant contribution to the changes in {\bf orientational} dynamics. This also allows us to rationalise the numerical  results~\cite{Popescu2009cavity,Popescu2014numerics}.

%
%
While hydrodynamics enhances the drag experienced by the swimmer, the wall-induced-diffusiophoresis enhances propulsion along the wall - in addition to perpendicular attraction (repulsion) for swimmers moving with their active-face-forward (inert-face-forward). The sign of the product of the swimmer mobility coefficient (determined by how solutes interact with the swimmer surface) and its net production rate ($\pm$ for source/sink) determines if it has its inert or active face forward.
%
%
%
%
\section{The model}
\begin{figure}
\begin{center}
\includegraphics[scale=0.6]{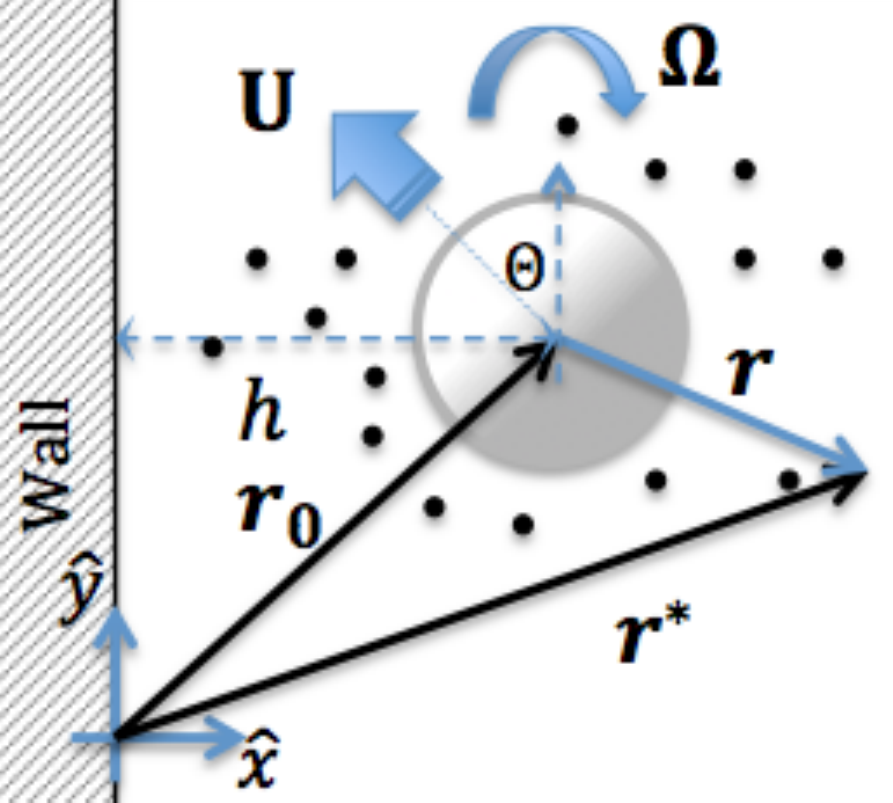}
\caption{(colour online) Schematic swimmer-wall problem. Black dots represents the solute molecules. The swimmer rotates with an angular velocity $\bs{\Omega}$ and translates with velocity $\mathbf{U}$ at a pitch angle $\Theta$ to the wall. } \label{swimmer_wall}
\end{center}
\end{figure}

We consider  self-phoretic Janus swimmers in the limit of vanishing P\'eclet (${\cal P}e$) and Reynolds (${\cal R}e$) numbers. Our goal is to obtain the swimmer propulsion speed $U$ as a function of the system parameters, such as  $D$, the solute diffusion coefficent, $R$, the swimmer radius and $h$, its  distance from the wall. A brief consideration of typical scales is useful at this point.
Janus swimmers~\cite{howse2007self,ebbens2014electrokinetic} with sizes $R= 1-2 \mu$m and speed $U = 1-10 \mu$m/s in a solution will have a P\'eclet number in the range of $\mathcal{P}e = UR/D \sim 10^{-3} - 10^{-2}$, where $D \sim 10^{-9}$m$^2$s$^{-1}$ is the solute diffusion coefficient. This implies to leading order, 
advection of solute particles by the flow generated by the swimmer is negligible compared to their diffusion \cite{khair2013peclet}. A useful interpretation of the Pecl\'et number is provided by the comparison of two timescales $\mathcal{P}e = \tau_D/\tau_P$;  the diffusive time-scale $(\tau_D \sim R^2/D)$ of the fuel solutes and the swimmer propulsive time-scale  $(\tau_P \sim R/U)$. 
 Inertia also plays a negligible role (as $\mathcal{R}e = UR/\nu\sim 10^{-6} \ll 1$ for typical  
 solution kinematic viscosity $\nu\sim 10^{-6}$m$^2$/s). 
 Hence the solute concentration profile is in steady state with the bulk and  satisfies the Laplace equation
\begin{equation}
\nabla^2 C ( \bs{r})  = 0 \label{solute:diffusion},
\end{equation}
where $\bs{r} = \bs{r}^* - \bs{r}_0 \equiv (x,y,z)$ with $\bs{r}_0$, the position of the swimmer centre.
The catalytic chemical reaction happening on the surface of the swimmer generates a radial flux $\alpha(\bs{\hat{n}})$ (see fig. \ref{swimmer_wall}), which gives the boundary condition (BC)
\begin{equation}
- \left. D \bs{\hat{n}} \cdot \nabla C (\bs{r})\right|_{r=R}  = \alpha (\bs{\hat{n}}) \label{surface:chemical:activity}
\end{equation}
This flux condition will be be interpreted as the effective flux of the solutes at the outer edge of  an  ``interaction layer".   Examples of $\alpha(\bs{\hat{n}})$ are surface flux of half coated Janus swimmer in fuel solution in which $\alpha(\bs{\hat{n}}) = \alpha K(\bs{\hat{n}})$ in the reaction-limited regime and $\alpha(\bs{\hat{n}}) = \alpha C(\bs{\hat{n}}) K(\bs{\hat{n}})$ in the diffusion-limited regime. $K(\bs{\hat{n}})$ is 1 on the \textit{active} hemisphere and 0 on the \textit{inert} hemisphere. In the following we will study the catalytic kinetics in the reaction-limited regime; i.e the bulk fluid serves as fuel bath \cite{Golestanian01,Bechinger01}. 
Furthermore, we consider the wall to be inert and impermeable to the solutes
\begin{equation}
- \left. D \bs{\hat{x}} \cdot \nabla C(\bs{r}) \right|_{x=-h} = 0 \label{wall:impermeability}
\end{equation}
Far away from the wall and the swimmer surface, the concentration of the solute takes the bulk value $C \rightarrow C_{\infty}, \quad \{x\rightarrow + \infty, y,z \rightarrow \pm \infty \}$.\\
\par In the zero ${\cal R}e$ limit, the fluid flows $\bs{v}$ induced by the swimmer satisfy the Stokes eqn. (incompressible fluid) 
\begin{equation}
\eta \nabla^2 \bs{v}(\bs{r})  -  \nabla p(\bs{r}) = \mathbf{0}, \quad \nabla \cdot \bs{v}(\bs{r}) = 0 \label{stokes:and:continuity}
\end{equation}
in the half-plane (shown in fig. (\ref{swimmer_wall})) and $\eta$ is the solvent dynamic viscosity and $p$, hydrostatic pressure. The flow field satisfies the slip condition  
\begin{equation}
\left. \bs{v}(\bs{r}) \right|_{r=R} = \mathbf{U} + \bs{\Omega} \times \bs{r} + \bs{v}^s \label{sw:surface:noslip}
\end{equation}
on the swimmer surface, where $\mathbf{U}, \bs{\Omega}$ are the linear and angular velocities respectively - which are unknowns and the goals of this analysis. Phoretic slip $\bs{v}^s$ arises due to the viscous stresses balancing osmotic pressure gradient in the 'thin interaction region'~\cite{Anderson01}. The latter is generated by the coupled asymmetric distribution of the solutes $C(\bs{r})$ and their short-ranged interaction $\Psi(\bs{r})$ with the swimmer surface. The expression $\bs{v}^s = \mu \left( \mathds{1} - \bs{\hat{n}\hat{n}} \right) \cdot \nabla C$ is obtained by matching an  ``inner" (interaction layer) to the bulk fields, where $\mu =  \frac{\beta^{-1}}{\eta} \int_0^{\infty} \rho \left(1 - e^{- \beta \Psi (\rho) }  \right) d \rho$ is a mobility coefficient~\cite{Anderson01} and $\beta^{-1}=k_B T$. We also have the no-slip BC on the wall, $\bs{v}(\bs{r})|_{x=-h}=\mathbf{0}$ and vanishing hydrodynamic flow in the bulk, $ \bs{v} \rightarrow \mathbf{0} $, $\{x\rightarrow + \infty, y,z \rightarrow \pm \infty \}$. The swimmer have zero net body-force and torque.
\begin{equation}
 \oiint \bs{\Pi} \cdot \bs{\hat{n}} \ d \mathcal{S} = \mathbf{0}, \quad 
\oiint \bs{r} \times \left( \bs{\Pi} \cdot \bs{\hat{n}} \right) \ d \mathcal{S}  = \mathbf{0} \label{force:torque:constraint}
\end{equation}
where $ \bs{\Pi} = - p \mathds{1} + \eta \left( \nabla \bs{v}  + (\nabla \bs{v})^T \right)$ is the hydrodynamic stress tensor and $\mathds{1}$ is the unit tensor.
\section{Analysis} For the swimmer problem near the wall, we employ the so-called method of images (reflections) \cite{kimmicrohydrodynamics}. This involves first, finding the swimmer propulsion velocity in the bulk, far enough away from any boundaries. Subsequently, we use the bulk solution to find corrections to the pertinent fields; and hence finding the translational and angular velocities corrections due to the wall. 
\subsection{Swimming in the bulk (no wall)} In the absence of the wall ($h \rightarrow \infty$), the concentration field BCs becomes $C \rightarrow C^{\infty}, \ r \rightarrow \infty$, in addition to the swimmer surface flux BC (\ref{surface:chemical:activity}). The solute concentration field follows from (\ref{solute:diffusion}) for axisymmetric coating of the catalyst with symmetry axis $\bs{\hat{u}}$ \footnote{The advantage of writing the solution in this form is that calculation of the image system singularities for a swimmer with a pitch angle $\Theta$ relative to the wall becomes straight forward - rotating the symmetry axis $\bs{\hat{u}} \rightarrow \mathcal{R}(\Theta) \bs{\hat{u}}$ (where $\mathcal{R}\left(\Theta\right)$ is the rotation matrix).}
\begin{equation}
C^{(0)} = C^{\infty} + \frac{R}{D} \sum_{k=0}^{\infty} \frac{1}{k+1} \alpha^{(k)} P_k(\bs{\hat{u} \cdot \hat{r}}) \left( \frac{R}{r} \right)^{k+1} \label{c:general:solution}
\end{equation}
where $P_k(\bs{\hat{u} \cdot \hat{r}})$ are the normalised Legendre polynomials such that 
\begin{math}
P_0 = 1, \quad P_1 = \bs{\hat{u} \cdot \hat{r}}, \quad P_2 =  \bs{\hat{u}}\bs{\hat{u}} : \left( 3 \bs{\hat{r}\hat{r}} - \mathds{1}\right)/2 \; , 
\end{math}
and $\alpha^{(k)}$ are the surface moments of the catalytic solute flux $\alpha(\bs{\hat{n}})$ (see equation \ref{surface:chemical:activity}):
\begin{math}
\alpha^{(k)}  = \left( k + \frac{1}{2} \right) \int_{-1}^1 d \left( \bs{\hat{u} \cdot \hat{r}} \right) \alpha \left( \bs{\hat{r}} \right) P_k \left( \bs{\hat{u} \cdot \hat{r}}\right) \, . 
\end{math}
Now, substituting the solute concentration field $C^{(0)}$ into the slip velocity $\bs{v}^s= \mu \left( \mathds{1} - \bs{\hat{n} \hat{n}}\right) \cdot \nabla C^{(0)}$ gives
\begin{align}
\bs{v}^s & = \frac{\mu \alpha^{(1)}}{3D} \bs{\hat{u}}  +   \frac{\mu \alpha^{(1)}}{6D}\left( \mathds{1} - 3 \bs{\hat{n} \hat{n}} \right)\cdot \bs{\hat{u}}  + \mathcal{O}\left( \alpha^{(2)} \right) \label{slip:velocity}
\end{align}
\par To simplify our presentation, we consider a self-phoretic swimmer with linear catalytic coating $\alpha(\bs{\hat{n}}) = \alpha^{(0)} + {\alpha}^{(1)} \bs{\hat{n}}\cdot \bs{\hat{u}}$. (setting $\alpha^{(k)} = 0$ for $k \geq 2$;  including e.g. $\alpha^{(2)}$ does not qualitatively change the types of behaviour observed).
Therefore, the unbounded domain solute concentration from eqn. (\ref{c:general:solution}) gives (see Fig. \ref{c:field:isolated}) 
 \begin{align}
C^{(0)} (\bs{r}) & = C^{\infty}  + \frac{\alpha^{(0)} R }{D} \underbrace{ \left( \frac{R}{r} \right) }_{ \oplus} + \frac{\alpha^{(1)} R }{2 D} \underbrace{  \left(\frac{R}{r}\right)^2 \bs{\hat{u}} \cdot \bs{\hat{r}} }_{ \ominus\oplus}  \label{c:field:isolated}
 \end{align}
\begin{figure}[h!]
\includegraphics[scale=0.43]{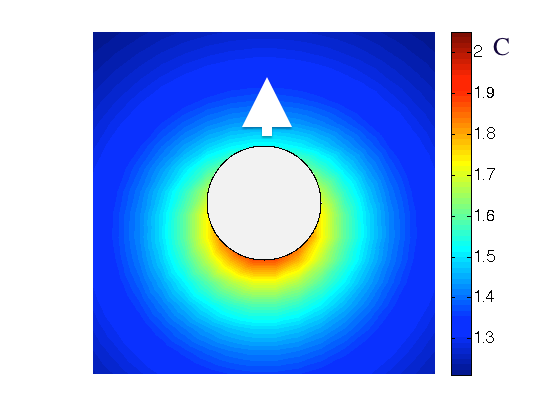}
\caption{Solute density profile for a  \textit{source}-swimmer $\alpha^{(0)} > 0$ in the bulk (from eqn. (\ref{c:field:isolated})).}
\end{figure}
The fluid flow field satisfy the isotropic vanishing flow field in the bulk, $ \bs{v} \rightarrow \mathbf{0}, \ r \rightarrow \infty$ and from equations (\ref{stokes:and:continuity},\ref{slip:velocity}), the flow field results \cite{Anderson01} :
 \begin{math}
\bs{v}^{(0)} (\bs{r})=   \frac{1}{2}\left( \frac{R}{r} \right)^3 \left ( 3 \frac{\bs{r r}}{r^2} - \mathds{1} \right ) \cdot  \mathbf{U}_0 \label{v:field:isolated}
\end{math}
with zero hydrostatic pressure gradient $p = p_{\infty}$, where the propulsion $\mathbf{U}_0$ and the angular velocities $\bs{\Omega}_0$ are obtained using the force and torque balance constraints (\ref{force:torque:constraint}) as
\begin{equation} 
 \mathbf{U}_0 = - \frac{ \mu \alpha^{(1)}}{3 D} \bs{\hat{u}}, \qquad \bs{\Omega}_0 = \mathbf{0}  \label{propulsion:vel} \, . 
\end{equation}
We have restricted ourselves to  a uniform mobility $\mu$ over the swimmer surface, leading to zero rotation : $\bs{\Omega}_0 = \mathbf{0}$.  We note that $\alpha^{(1)}$ and the symmetry direction $\bs{\hat{u}}$ are not  independent: choosing $\bs{\hat{u}}$ fixes the sign of $\alpha^{(1)}$.
\begin{figure}
\begin{center}
\includegraphics[scale=.4]{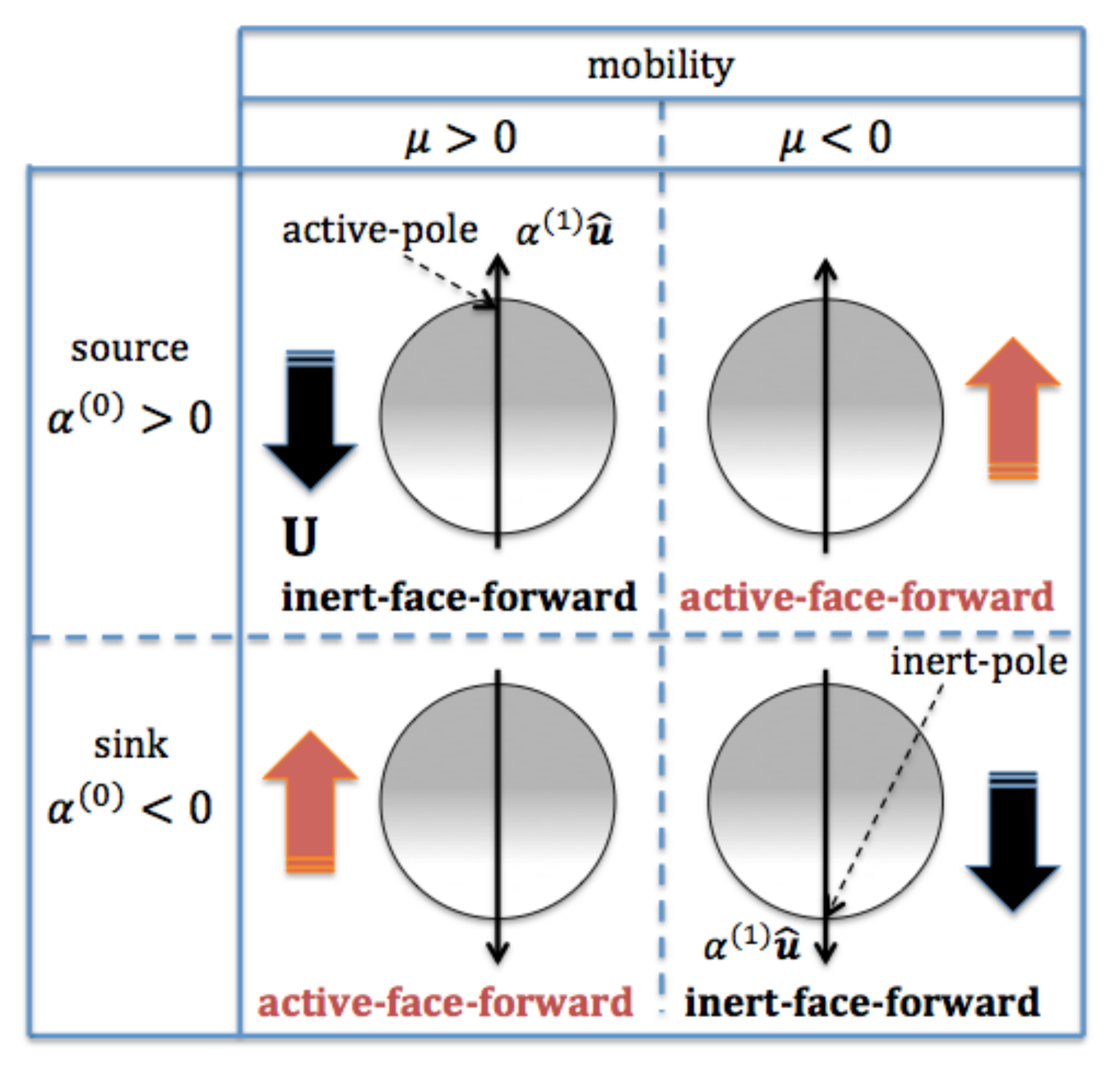}
\caption{Swimming direction for different combinations of the mobility $\mu$ (containing the solute-to-surface interaction information) and swimmer type $\alpha^{(0)}$.} \label{swimming_direction}
\end{center}
\end{figure}

\subsection{Swimming near a wall} We now consider a swimmer whose centre is a distance $h$ from an infinite plane wall (see Fig. \ref{swimmer_wall}).
We proceed by finding corrections to the bulk velocities $\left( \begin{array}{c}\mathbf{U} \\ \bs{\Omega} \end{array}\right) =\left( \begin{array}{c} \mathbf{U}_0+\mathbf{U}_1 + \ldots \\ \bs{\Omega}_0+ \bs{\Omega}_1 + \ldots \end{array}\right)$. This is achieved by adding singular flow and concentration fields $(\bs{v}^{(1)}(\bs{r}),C^{(1)}(\bs{r}))$ centred behind the wall (at the image point) to impose the the no-slip and the impermeability conditions on the wall. Furthermore since adding them means the flow no longer satisfies the BCs on the swimmer surface, we add further singular fields $(\bs{v}^{(2)}(\bs{r}),C^{(2)}(\bs{r}))$, this time centred at the swimmer centre to maintain the correct slip and constant flux BCs. This process can be iterated yielding to a power series solution in $\epsilon=R/h$. 
\begin{figure}[h!]
\begin{center}
\includegraphics[scale=0.4]{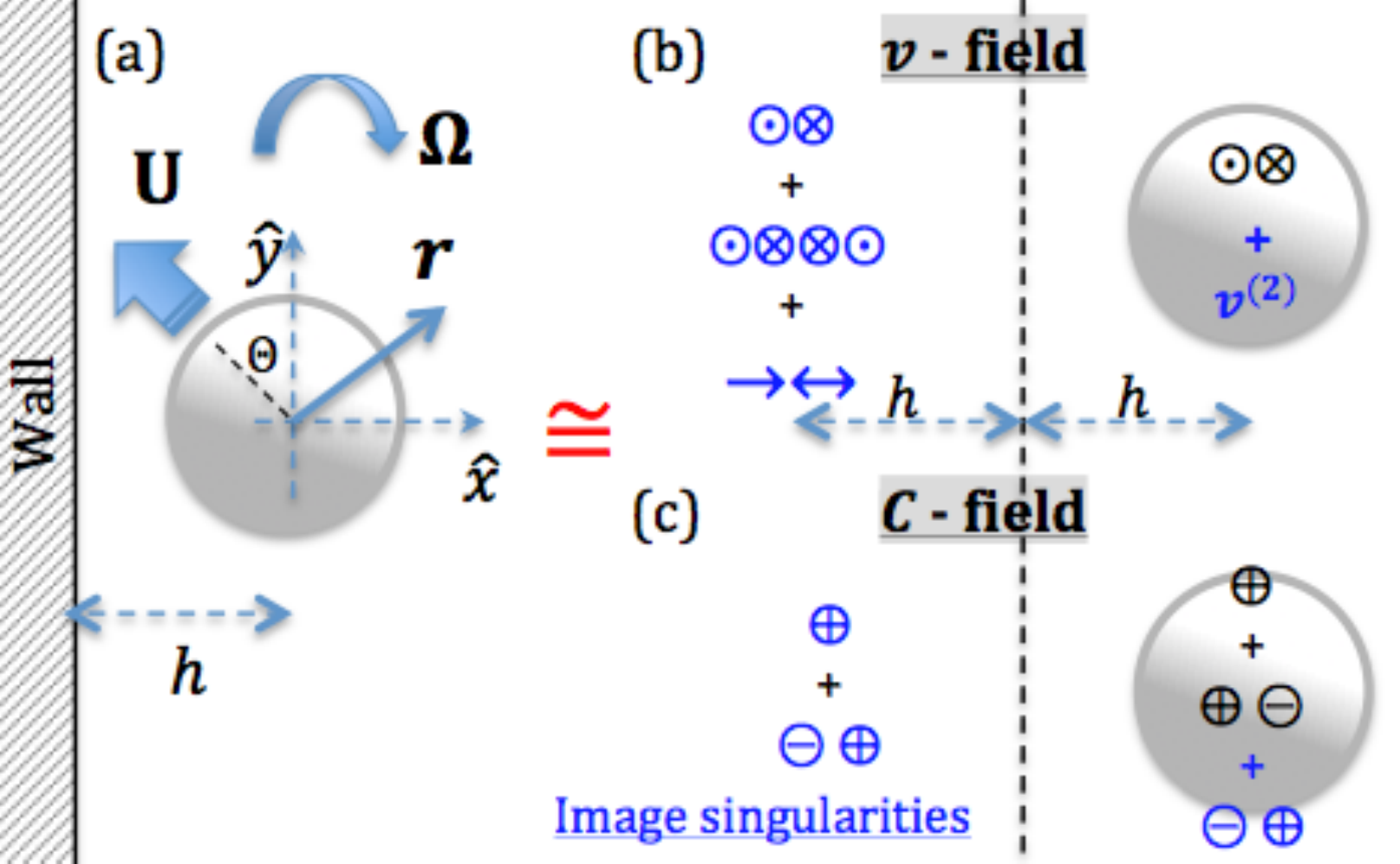}
\caption{
(colour online) Flow velocity and solute concentration fields for swimmer moving near a wall. The image system of the flow field is (source-doublet \textcolor{blue}{$\otimes\odot$} + source-quadrupole \textcolor{blue}{$\odot\otimes\otimes\odot$} + force-quadrupole \textcolor{blue}{$\rightarrow\leftrightarrow$}) \cite{blake,Lauga} while that of the concentration field is (solute-monopole \textcolor{blue}{$\oplus$} + solute-dipole \textcolor{blue}{$\ominus\oplus$}). See equations (\ref{c1:image}, \ref{c2:image} and \ref{v1:image}). 
} 
\label{swimmer:wall:image}
\end{center}
\end{figure}
Here we keep only the leading order terms :
\begin{align}
\bs{v}(\bs{r}) & = \bs{v}^{(0)} + \bs{v}^{(1)} + \bs{v}^{(2)} + \mathcal{O}\left(\left[R/h \right]^6\right) \\
C (\bs{r}) & = C^{(0)} + C^{(1)} + C^{(2)} + \mathcal{O}\left(\left[R/h\right]^3\right) \label{c:approx}
\end{align}
The wall reflected concentration field is 
\begin{equation}
C^{(1)} = \frac{\alpha^{(0)} R}{D}  \left( \frac{R}{r'} \right) + \frac{\alpha^{(1)} R}{2D}  \left( \frac{R}{r'} \right)^2 \bs{\hat{r}}' \cdot \left( \bs{\hat{u}}^{\parallel} - \bs{\hat{u}}^{\perp}\right)  \label{c1:image}
\end{equation}
where $\bs{r}' = \bs{r} + 2h \bs{\hat{x}}$, $\bs{\hat{u}}^{\parallel} =  \cos \Theta \bs{\hat{y}}$  and $\bs{\hat{u}}^{\perp} = - \sin \Theta \bs{\hat{x}}$, while the image singularities $\bs{v}^{(1)}$ for bulk flow $\bs{v}^{(0)}$ 
are well known (see Appendix A,\cite{blake,Lauga}). Furthermore, the swimmer surface reflected solute concentration field is given by
\begin{equation}
C^{(2)}  =  \frac{\alpha^{(0)} R^3}{32D r^2} \left[ -4\epsilon^2  \bs{\hat{x}} +  \frac{\alpha^{(1)} \epsilon^3  }{\alpha^{(0)}}    \left( \bs{\hat{u}^{\parallel}} + 2 \bs{\hat{u}^{\perp}} \right)  \right] \cdot \bs{\hat{r}} \label{c2:image}
\end{equation}
and since we are only interested in the leading order rigid body corrections $\left( \mathbf{U}_1, \bs{\Omega}_1 \right)$\cite{diffcomment}, we can bypass finding an explicit expression for $\bs{v}^{(2)}$ using the reciprocal theorem \cite{kimmicrohydrodynamics}.
\subsection{Wall-induced-diffusiophoresis}
The wall distortion on the solute concentration field is approximated by images made of a monopole and dipoles (see fig. \ref{swimmer:wall:image}(c) and eqns. \ref{c1:image},\ref{c2:image}). These reflected fields do not induce rotation $\bs{\Omega}^d_1 = \mathbf{0}$, but enhance the gradients giving the linear translation wall-induced-diffusiophoresis 
\begin{equation} 
\mathbf{U}^d_1 = \frac{\mu \alpha^{(0)}}{4D} \epsilon^2   \bs{\hat{x}} + \frac{3}{16} \left( \mathbf{U}_0^{\parallel} + 2\mathbf{U}_0^{\perp} \right) \epsilon^3 + \mathcal{O} \left( \epsilon^4 \right) \label{w:i:d:wall}
\end{equation}
We note that the leading order term  $(\sim \epsilon^2)$ is present irrespective of the swimmer orientation. Its strength is determined by the net consumption/production rate: $\alpha^{(0)}$. 
This may be contrasted with an orientation dependent  leading order force-dipole contribution in squirmer models~\cite{Lauga,Ishimoto}. 
In effect, $\mathbf{U}^d_1$ repels (attracts) inert-face-forward (active-face-forward) swimmers to (from) the wall, while enhancing their parallel propulsion along the wall.
\begin{figure}
\includegraphics[scale=0.43]{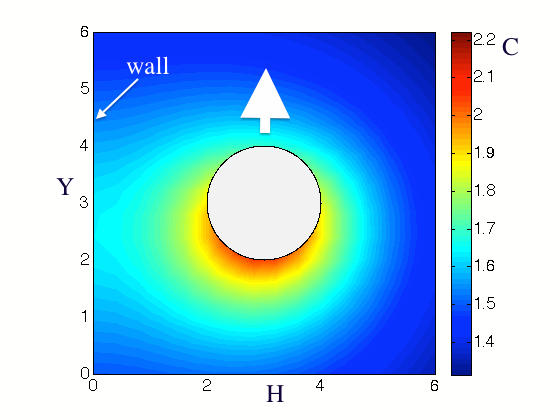}
\caption{Solute density profile for a \textit{source}-swimmer $\alpha^{(0)} > 0$ near a wall from eqn. (\ref{c:approx}).} 
\includegraphics[scale=0.33]{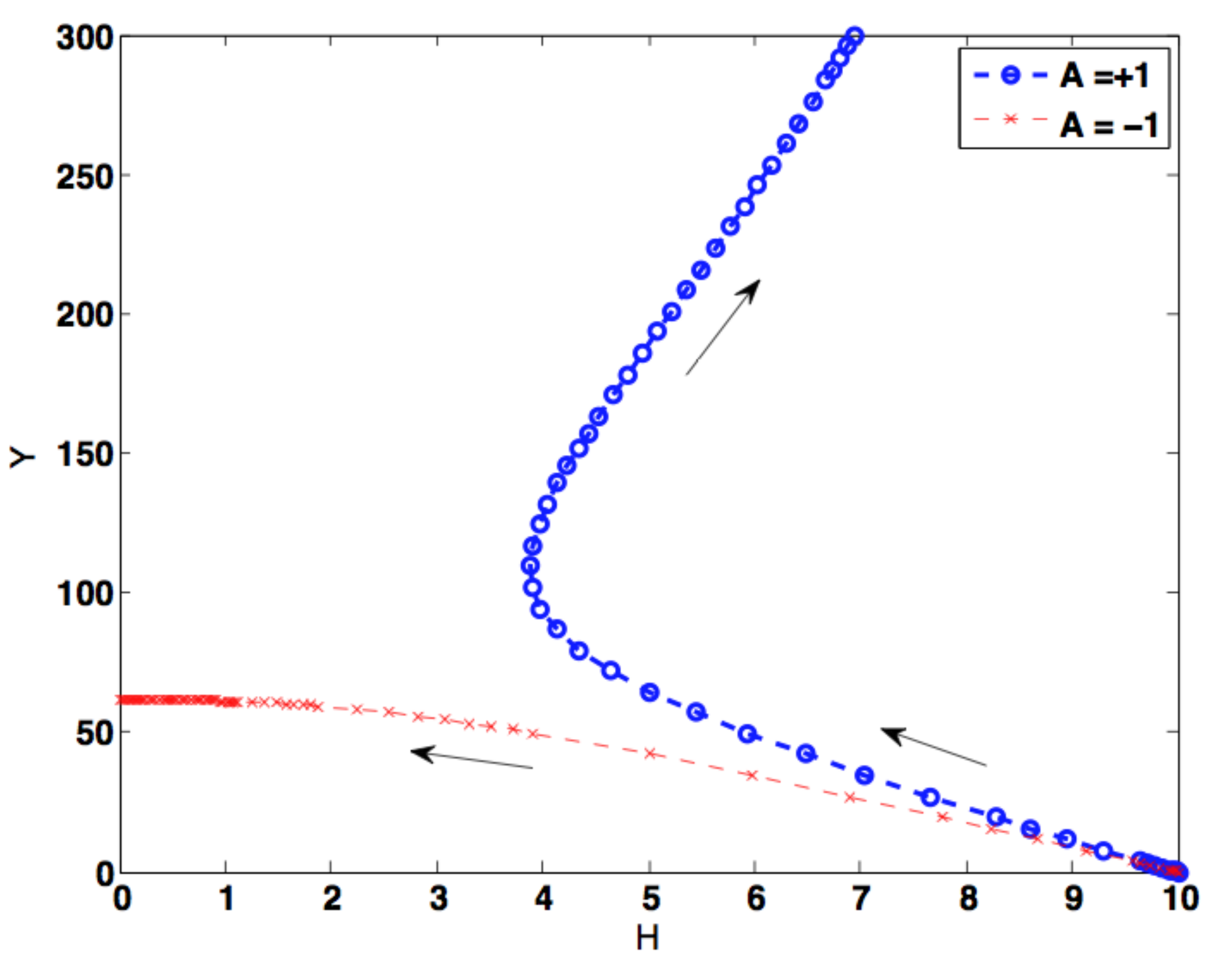}
\caption{(color online) Typical swimmer trajectory for initial condition $(h(0),y(0),\Theta(0))=(10R,0,0.1)$ and $A = \pm 1$. Trajectories of an \textit{active-face-forward} swimmer $(A=-1)$ with initial orientations facing the wall are attracted to the wall.
} 
\label{trajectory:sample}
\end{figure}

\subsection{Hydrodynamic contribution}
The no-slip condition on the wall introduces the reflected image field $\bs{v}^{(1)}$ (\ref{v1:image}) (see also ref. \cite{blake,Lauga}). This contribution enhances the drag experienced by the swimmer \cite{Anderson1985boundary,Lauga}
\begin{equation}
\mathbf{U}^h_1 = - \frac{\epsilon^3}{8} \left( \mathbf{U}^{\parallel}_0 + 4 \mathbf{U}^{\perp}_0 \right) + \mathcal{O}\left( \epsilon^6 \right) \label{hydro:drag:wall}
\end{equation}
In addition, the wall induces rotation with the swimmer re-orienting weakly with angular velocity \cite{Anderson1985boundary,Lauga}
\begin{equation}
\bs{\Omega}^h_1  =   \frac{3}{16 R} \epsilon^4 \ \mathbf{U}^{\parallel}_0 \times \bs{\hat{x}} + \mathcal{O}\left( \epsilon^7\right) \label{reorientation}
\end{equation}
Equations (\ref{w:i:d:wall}), (\ref{hydro:drag:wall}) and (\ref{reorientation})  imply that if orientation  fluctuations are ignored, the swimmer trajectory is confined to the $x-y$ plane.

\subsection{Dynamical system}
The results described above  can be summarised by a dynamical system for the position of the swimmer centre, $\bs{r}_0(t) \equiv (h(t),y_0(t),z_0(t))$ in the laboratory frame, placing the origin on the wall.  The swimmer will trace a trajectory following the kinematic equations
\begin{equation}
\frac{d \bs{r}_0}{d t} (t) = \mathbf{U} (t); \quad
\frac{d \bs{\hat{u}}}{d t} (t) = \bs{\Omega} \times \bs{\hat{u}} (t) 
\end{equation}
where $\mathbf{U} = \mathbf{U}_0 + \mathbf{U}_1^d + \mathbf{U}_1^h$ and $\bs{\Omega} = \bs{\Omega}_1^h$. 
Non-dimensionalising using, $H(t)={h(t) \over R}, Y(t)={ y_0(t) \over R},Z(t)={z_0(t)\over R}$, $\tau={t U_0 \over R}$, we obtain rescaled kinematics ($\dot H = {d H \over d \tau}, \dot Y = {d Z \over d \tau},\dot H = {d Z \over d \tau}$) for position :
\begin{equation} 
\begin{bmatrix}
\dot{H} \\
\dot{ Y} \\
\dot{Z} 
\end{bmatrix} =
\begin{bmatrix}
-\sin \Theta  \\
\cos \Theta \\
0 
\end{bmatrix} + \frac{A}{H^2} \begin{bmatrix}
1\\0\\0
\end{bmatrix} + \frac{1}{16H^3}\begin{bmatrix}
 2\sin \Theta  \\
\cos \Theta \\
0 
\end{bmatrix}\label{nonlinear:odes:1}
\end{equation}
and  orientation ($\dot \Theta = {d \Theta \over d \tau}$)
\begin{equation}
\dot{\Theta} = - \frac{3}{16H^{4}} \cos \Theta 
\end{equation}
with a dimensionless parameter  $A = (\mu \alpha^{(0)}/4D)/U_0$, where  $U_0 = |\mu \alpha^{(1)}/3D|$ is the bulk speed. 

We can therefore classify the behaviour of this dynamical system by the sign of $A$. We note that 
\textit{active-face-forward} swimmers have $(A < 0)$ while \textit{inert-face-forward} swimmers have $(A>0)$ (see fig. (\ref{swimming_direction})). 
Inert-face-forward swimmers are always repelled while active-face-forward ones are always attracted to the wall irrespective of their orientation. We emphasise however, the eventually both escape from the vicinity of the wall  with swimming directions oriented away from the wall.
%
%

The swimmer dynamical system (\ref{nonlinear:odes:1}) has two steady modes of motion (one linearly stable and one linearly unstable). The unstable stationary state,  
$(\dot{H},\dot{Y},\dot{Z},\dot{\Theta})=\mathbf{0}$, which corresponds to a swimmer at a wall separation $\{  H_*: 8 H_*^3 - 8 |A| H_* = 1, H_* > 1 \}$ facing directly away from (or towards) the wall $\Theta_* = \begin{Bmatrix}
3\pi/2, \ \text{for} \ A< 0\\ \pi/2, \ \text{for} \ A > 0
\end{Bmatrix}$. 
The second (stable) mode of steady motion is the limiting case of the swimmer far away from the wall ${H\rightarrow \infty}$ with swimming orientation away from the wall $\Theta \in ( \pi, 2\pi)$ so that eventually the swimmer recovers its bulk propulsion behaviour. 

\par Numerically integrating the non-linear system (\ref{nonlinear:odes:1}) gives the swimmer trajectories near the wall (see fig. (\ref{trajectory:sample}) for a typical trajectory). 
We identify a turning point at $H_T$ at which a swimmer changes direction from pointing towards to pointing away from the wall. 
We note however, that for swimmer parameter values of $|A| > 1/4$, the wall separation for the \textit{active-face-forward} $H_T < 1$; which implies the swimmer will eventually crash into the wall (see fig. (\ref{trajectory:sample}) for a sample trajectory), noting of course however that the current analysis here is no longer valid when the swimmer gets too close to the wall where different physics governed by fluid incompressibility will dominate.

%
%
\section{Conclusion and discussion} In summary, we have  studied  
the dynamics of a self-diffusiophoretic spherical Janus swimmer moving near a planar wall. In our analysis, we have been able to separate the contribution of different 
mechanisms to the propulsion allowing us to identify the differences between these swimmers driven by chemical (phoretic) processes to traditionally studied swimmers driven by mechanical deformations.
We have obtained leading order contributions to 
 the wall-induced distortion of the solute concentration gradient and 
shown that the wall impermeability to the solutes introduces a new contribution to swimmer propulsion which we have called  \textit{wall induced-diffusiophoresis}. We emphasise that this is quite different from the effect of walls on hydrodynamic interactions~\cite{Ishimoto,Lauga}  as studied e.g. using the squirmer model. Further, we find a natural way to categorise Janus swimmers into two classes: (1) \textit{inert-face-forward} swimmers which  have an enhanced parallel propulsion along the wall before being scattered away due to a combination of hydrodynamic repulsion and wall-induced-diffusiophoresis and (2)   \textit{active-face-forward} swimmers which are strongly attracted to the wall.  
%
\par
The wall-induced-diffusiophoresis leads to migration either towards or away from the walls depending on whether the swimmer is a global source or sink.
Interestingly we note that wall-induced-diffusiophoresis is present even for symmetrically coated active colloids.
 This robust effect may play a role in the attraction of  phoretic swimmers to surfaces~\cite{Palacci}. 
 
\par Recently it has been observed that Platinum-Polystyrene Janus particles are propelled by electrochemical as well as concentration gradients \cite{brown2014ionic,ebbens2014electrokinetic}. 
Similar types of behaviour would be observed such a self-electrophoretic swimmer, because the leading order electric potential for these swimmers is a dipole and hence can also be dominated by   the leading order interaction due to solute concentration presented here. 
%
\par The following key assumptions were made in our analysis: the swimmer separation from the wall is large compared to the swimmer size such that higher order reflected fields from the wall and swimmer-surface be neglected~\cite{Lauga}.  In addition, we assume the concentration profile is at quasi-steady state, the catalytic flux on the swimmer surface is constant and we have also ignored orientational  fluctuations of the swimmers. It would thus be interesting in the future  to examine the dynamics for swimmers very close to the wall using e.g. a lubrication analysis and the role of fluctuations.

\acknowledgments
YI acknowledges the support of University of Bristol.

\section{Appendix A: Flow image system}
The image system for the source-doublet $\bs{v}^{(0)}$ \cite{blake,Lauga} is
\begin{align}
\bs{v}^{(1)}& = \overbrace{\mathds{D}(\bs{r}')}^{\textcolor{blue}{\otimes\odot}} \cdot \left( \mathbf{U}_0^{\parallel} - 3 \mathbf{U}_0^{\perp}\right) - 2h \overbrace{\partial_{x_0} \mathds{D}(\bs{r}')}^{\textcolor{blue}{\odot\otimes\otimes\odot}}\cdot \left( \mathbf{U}_0^{\parallel} -  \mathbf{U}_0^{\perp}\right) \nonumber \\
& \qquad +  \underbrace{\partial^2_{x_0y_0} \mathds{G}(\bs{r}')}_{\textcolor{blue}{\rightarrow\leftrightarrow}} \cdot \mathbf{U}_0^{\parallel} -  \underbrace{\partial^2_{x_0} \mathds{G}(\bs{r}')}_{\textcolor{blue}{\rightarrow\leftrightarrow}} \cdot \mathbf{U}_0^{\perp} \tag{A1} \label{v1:image}
\end{align}
where $\bs{r}' = \bs{r}+2h\bs{\hat{x}}$ and $\bs{r} = \bs{r}^* - \bs{r}_0$; $\partial_{x_0} \equiv \bs{\hat{x}} \cdot \nabla_0 $ and $\partial^2_{x_0y_0} \equiv \left( \bs{\hat{x}} \cdot \nabla_0 \right) \left( \bs{\hat{y}} \cdot \nabla_0 \right)$; $\mathbf{U}^{\parallel} =   \mathbf{U} \cdot \bs{\hat{y}}$ and $\mathbf{U}^{\perp} =   \mathbf{U} \cdot \bs{\hat{x}}$.
\begin{equation*}
\mathds{G}(\bs{r}') = \left(\frac{R}{r'}\right) \left(\mathds{1} + \bs{\hat{r}' \hat{r}'} \right); \quad \mathds{D}(\bs{r}') = \frac{1}{2} \left(\frac{R}{r'}\right)^3 \left( 3\bs{\hat{r}' \hat{r}'} - \mathds{1}  \right)
\end{equation*}
are the stokeslet and source-doublet tensors.
\section{Appendix B: Fax\'ens Theorems}
Due to the finite size of the swimmer, the reflected flow and concentration fields, $\bs{v}^{(1)}$ and $C^{(1)}$ do not satisfy the BC on the swimmer surface. Therefore, to impose the correct BC, we add $\bs{v}^{(2)}$ and $C^{(2)}$ at the swimmer centre for the flow and concentration fields respectively such that $\bs{\hat{n}} \cdot \nabla C^{(2)} = - \bs{\hat{n}} \cdot \nabla C^{(1)} $ and
\begin{equation}
\bs{v}^{(2)} = -\bs{v}^{(1)} + \mathbf{U}_1 + \bs{\Omega}_1 \times \bs{r} + \mu \nabla_s \left[ C^{(1)} + C^{(2)} \right] \tag{B1}
\end{equation}
are satisfied on the swimmer surface and $\nabla_s = (\mathds{1} - \bs{\hat{n} \hat{n}}) \cdot \nabla$. Therefore, applying the reciprocal theorem (\cite{kimmicrohydrodynamics})
\begin{equation}
\oiint \bs{v'} \cdot \left( \bs{\Pi}^{(2)} \cdot \bs{\hat{n}} \right) d \mathcal{S} = \oiint \bs{v}^{(2)} \cdot \left( \bs{\Pi'} \cdot \bs{\hat{n}} \right) d \mathcal{S} \label{reciprocal:theorem} \tag{B2}
\end{equation}
with $\bs{\Pi}=  - p \mathds{1} + \eta \left( \nabla \bs{v} + \left[ \nabla \bs{v} \right]^T \right) $ and $\bs{v'}$ been an arbitrary external stokes flow which satisfies $\bs{v'} = \mathbf{U'}$ on a sphere of radius $R$ and $\bs{v'} \rightarrow \mathbf{0}$ for $r \rightarrow \infty$. Its well known that pure translating sphere have constant surface traction $\bs{\Pi'} \cdot \bs{\hat{n}} = - \frac{3 \eta}{2 R} \mathbf{U'}$ and hence
\begin{equation}
\mathbf{U'} \cdot \oiint \left(\bs{\Pi}^{(2)} \cdot \bs{\hat{n}} \right) \ d \mathcal{S} =  - \frac{3 \eta}{2 R} \mathbf{U'} \cdot \oiint \bs{v}^{(2)}(\bs{r}) \ d \mathcal{S} \tag{B3}  \label{force_free}
\end{equation}
The swimmer is force free, which implies $\oiint \bs{v}^{(2)}(\bs{r}) \ d \mathcal{S} = \mathbf{0}$. Hence,
Taylor expanding the $\bs{v}^{(1)}$ and $C^{(1)}$ fields; the leading order rigid body translation results
\begin{align} 
\mathbf{U}_1 & =  \overbrace{\bs{v}^{(1)} (\mathbf{0}) \ + \ \frac{R^2}{6} \nabla^2 \bs{v}^{(1)} (\mathbf{0})}^{\mbox{hydrodynamic contribution}} \nonumber \\  & \quad - \underbrace{\frac{2 \mu }{3} \nabla C^{(1)}(\mathbf{0})  - \frac{\mu}{4 \pi R^2} \oiint \nabla_s  C^{(2)}(\bs{r}) \ d \mathcal{S}}_{\mbox{wall-induced-diffusiophoresis}} \tag{B4} \label{translation:correction_b} 
\end{align}

Similarly, for arbitrary pure rotation $\bs{\Omega'}$ of a sphere of radius $R$ with $\bs{v'} = \bs{\Omega'} \times \bs{r}$; the reciprocal theorem reads
\begin{equation*}
\bs{\Omega'} \cdot \oiint \bs{r} \times \left( \bs{\Pi}^{(2)} \cdot \bs{\hat{n}} \right) d \mathcal{S} = -3 \eta \bs{\Omega'} \cdot \oiint \bs{\hat{n}} \times \bs{v}^{(2)}(\bs{r}) \ d \mathcal{S} 
\end{equation*}
with $\Pi' \cdot \bs{\hat{n}} = - 3 \eta \bs{\Omega'} \times \bs{\hat{n}}$ and also since the swimmer must remain torque free, it rotates with 
\begin{equation} 
\bs{\Omega}_1 = \frac{1}{2}  \nabla \times \bs{v}^{(1)} (\mathbf{0})  - \frac{3\mu}{8 \pi R^3} \oiint \bs{\hat{n}} \times \nabla_s C^{(2)}(\bs{r}) \ d \mathcal{S} \tag{B5}
\end{equation}
where the mobility $\mu$ is assumed to be uniform.



\begin{thebibliography}{0}
\bibitem{diffcomment}
There is an additional symmetric dipole term that does not contribute to $(\mathbf{U}_1,\bs{\Omega}_1)$.
\end{thebibliography}
\end{document}